\documentstyle[12pt]{article}

\newcommand{\cZ}{{\cal{Z}}}
\newcommand{\cD}{{\cal{D}}}
\newcommand{\cC}{{\cal{C}}}
\newcommand{\Z}{{Z \!\!\! Z}}
\newcommand{\beq}{\begin{equation}}
\newcommand{\eeq}{\end{equation}}
\newcommand{\beqn}{\begin{eqnarray}}
\newcommand{\eeqn}{\end{eqnarray}}
\newcommand{\CK}[1]{\mbox{\scriptsize c}_{\mbox{$\scriptstyle #1$}}}
\newcommand{\nsum}[2]{\sum_{ #1(\CK{#2}) \in \Z }}

\newcommand{\nddsum}[2]{\sum_{\stackrel{\scriptstyle \dual #1(\dual\CK{#2})
\in \Z} {\delta \dual #1=0}}}
\newcommand{\dd}{\mbox{d}}
\newcommand{\dual}{\mbox{}^{\ast}}
\newcommand{\bbbz}{{\mathchoice {\hbox{$\sf\textstyle Z\kern-0.4em Z$}}
{\hbox{$\sf\textstyle Z\kern-0.4em Z$}}
{\hbox{$\sf\scriptstyle Z\kern-0.3em Z$}}
{\hbox{$\sf\scriptscriptstyle Z\kern-0.2em Z$}}}}
\newcommand{\LL}{{I\!\! L}}
\newcommand{\intpi}{\int\limits_{-\pi}^{+\pi} {\cD}}
\newcommand{\intinf}{\int\limits_{-\infty}^{+\infty} {\cD}}
\newcommand{\expb}[1]{\exp\left\{ #1 \right\} }
\newcommand{\ie}{\hbox{\it i.e.}{}}
\newcommand{\etc}{\hbox{\it etc.}{}}
\newcommand{\eg}{\hbox{\it e.g.}{}}

\newcommand{\re}[1]{(\ref{#1})}
\newcommand{\half}{\frac 12}
\newcommand{\const}{\mbox{const.}\cdot}
\newcommand\Appendix[1]{\par
\setcounter{section}{0}
 \setcounter{equation}{0}
 \renewcommand{\thesection}{Appendix \Alph{section}}
\section{#1}
 \def\theequation{\Alph{section}.\arabic{equation}}}
\newcommand\appendixn[1]{\par
 \setcounter{equation}{0}
 \renewcommand{\thesection}{Appendix \Alph{section}}
\section{#1}
 \def\theequation{\Alph{section}.\arabic{equation}}}

\title{
String Representation \\ of the Abelian Higgs Theory\\
and Aharonov--Bohm Effect on the Lattice
}
\author{ M.I.~Polikarpov\\
{\sl ITEP, Moscow, 117259, Russia},\\
{\small e-mail: polykarp@vxdsyc.desy.de}\\
 U.-J.~Wiese\\
{\sl HLRZ J\"ulich, 5170 J\"ulich, Germany}\\
and M.A.~Zubkov\\
{\sl ITEP, Moscow, 117259, Russia}}

\date{}

\def\JP{ J.~Phys.}

\def\NP{ Nucl.~Phys.}
\def\ZP{ Z.~Phys.}

\def\PRL{ Phys.~Rev.~Lett.}

\begin{document}
\bibliographystyle{bibstand}
\begin{titlepage}
\maketitle
\begin{abstract}
       The partition function of the $4D$ lattice Abelian Higgs
theory is represented as the sum over world sheets of Nielsen--Olesen
strings. The creation and annihilation operators of the strings are
constructed. The topological long--range interaction of the
strings and charged particles is shown to exist; it is proportional to
the linking number of the string world sheet and particle world trajectory.

\end{abstract}

\mbox{}
\vspace{-89ex}
\centerline{ }
\centerline{\bf INSTITUTE OF THEORETICAL AND EXPERIMENTAL PHYSICS}
\mbox{}\\
\mbox{}\hfill {\bf ITEP-16-1993}

\vfill
\centerline{MOSCOW 1993}

\thispagestyle{empty}
\end{titlepage}
\pagebreak
\mbox{}
\vspace{2cm}
\section{Introduction}

        There are several examples in the lattice field theory showing that
a change of the variables in the functional integral allows for a
formulation of the theory in terms of the physical excitations, \eg\ the
partition function of the $2D$ $XY$-model is equivalent to the partition
function of the Coulomb gas \cite{Ber70,KoTh73}, the partition function of
the $4D$ compact electrodynamics can be represented as a sum over the
monopole--antimonopole world lines \cite{BaMeKo77}. In the present
publication we show that the partition function of the four--dimensional
Abelian Higgs theory can be represented as the sum over closed surfaces
which are the world sheets of the Abrikosov--Nielsen--Olesen strings
\cite{Abr57,NiOl73}. We construct the string creation operators, which
create the string world sheets spanned on the given loops. If the field
which is condensed has the charge $Ne$, then there is a non--trivial
long--range topological interaction of Nielsen--Olesen strings with
particles of charge $Me$, provided that $\frac MN$ is non--integer. This
long--range interaction is described by the term in the action which is
proportional to the linking number of the string world sheet and the world
line of charge $Me$. This is the four--dimensional analogue
\cite{AlWi89,AlMaWi90,PrKr90} of the Aharonov--Bohm effect: strings
correspond to solenoids which scatter charged particles.

\section{World Sheets of the Nielsen--Olesen Strings}

        We consider the model describing interaction of the
noncompact gauge field $A_\mu$ with the scalar field $\Phi =
|\Phi|e^{i\varphi}$, whose charge is $Ne$. The selfinteraction of the
scalar field is described by the potential $V = \lambda (|\Phi|^2 -
\eta^2)^2$. For simplicity, we consider the limit as $\lambda \rightarrow
\infty$, so that the radial part of the scalar field is frozen, and the
dynamical variable is compact: $\varphi \in (-\pi,\pi]$.  The partition
function for the Villain form of the action is given by:

\beq
 \cZ = \intinf A \intpi \varphi \nsum{l}{1}
         \expb{ -S_l(A,\dd\varphi)}, \label{ZNC}
\eeq
where
\beq
 S_l(A,\dd \varphi) = \frac{1}{2e^2}\|\dd A\|^2
        + \frac{\kappa}{2} \| \dd\varphi + 2\pi l - N A \|^2 .
\label{ZNCS}
\eeq
We use the notations of the calculus of differential forms on the lattice
\cite{BeJo82}, which are briefly described in Appendix A. The symbol
$\int\cD\varphi$ ($\int\cD A$) denotes the integral over all site (link)
variables $\varphi$ ($A$). Fixing the gauge $\dd\varphi = 0$, we get the
following expression for the action \re{ZNCS}:  $S_l =
\frac{1}{2e^2}[A,(\delta\dd + N^2 \kappa e^2) A] + \ (terms \ linear \ in \
A)$; therefore, due to the Higgs mechanism, the gauge field acquires the
mass $m = N \kappa^\half e$; there are also soliton sectors of the Hilbert
space which contain Abrikosov--Nielsen--Olesen strings, hidden in
the summation variable $l$ in \re{ZNC}.

        The partition function of the compact electrodynamics can be
represented as a sum over closed world lines of monopoles
\cite{BaMeKo77}. In the same way the partition function \re{ZNC} can be
rewritten as the sum over closed world sheets of the Nielsen--Olesen
strings\footnote{We use the superscript BKT,
since a similar transformation of the partition function was first found
by Berezinskii \cite{Ber70}, Kostrlitz and Thouless \cite{KoTh73}, who
showed that the $XY$ model is equivalent to the Coulomb gas in two
dimensions}:

\beq
\cZ ^{BKT}= \const \nddsum{\sigma}{2} \expb{- 2\pi^2\kappa (\dual
\sigma,(\Delta + m^2)^{-1}\dual \sigma)}. \label{ZNCBKT}
\eeq
The derivation of this representation is given in Appendix B.
The sum here is over the integer variables $\dual \sigma$, which are
attached to the plaquettes $\dual c_2$ of the dual lattice. The condition
$\delta \dual \sigma = 0$ means that for each link of the dual lattice the
``conservation law'' is satisfied: $\sigma_1 + \sigma_2 + \sigma_3 =
\sigma_4 + \sigma_5 + \sigma_6$, where $\sigma_i$ are integers
corresponding to plaquettes connected to the considered link. The signs of
$\sigma_i$'s in this ``conservation law'' are dictated by the definition of
$\delta$ (by the orientation of the plaquettes 1,...,6). If $\sigma=0,1$,
then the condition $\delta \dual \sigma = 0$ means that we consider closed
surfaces made of plaquettes with $\sigma = 1$. In \re{ZNCBKT} we have
$\sigma \in \Z$, which means that one plaquette may ``decay'' into several
ones, but still the surfaces, made of plaquettes with $\sigma \neq 0$, are
closed. It follows from \re{ZNCBKT} that the strings interact with each
other via the Yukawa forces\footnote{Due to the definition of the
integration by parts $(\varphi,\delta\psi) = (\dd\varphi,\psi)$, the
operator \mbox{$(\Delta + m^2)^{-1}$} (and not \mbox{$(-\Delta +
m^2)^{-1}$}) is positively defined on the Euclidean lattice} -- $(\Delta +
m^2)^{-1}$.

\section{String Creation Operators}

        The creation of a string (as a nonlocal object) involves
nonlocal operators. Strings are surrounded by a
cloud of bosons, just as charged particles are surrounded
by their photon cloud. Creation operators for charged particles were first
constructed by Dirac \cite{Dir55}, whose idea was to compensate the gauge
variation of a charged field $\Phi(x)' = \Phi(x) \exp(i \alpha(x))$ by a
contribution of the gauge field representing the photon cloud:

\beq
\Phi_c(x) = \Phi(x) \expb{i \int
d^3 y B_i(x - y) A_i(y)}, \label{Dirac}
\eeq
where $\partial_i B_i(x) = \delta(x)$, and $A_i(x)' = A_i(x) + \partial_i
\alpha(x)$ is the photon field. The gauge invariant operator $\Phi_c(x)$
creates a scalar charged particle at point $x$, together with the photon
cloud surrounding it. Our construction of string creation operators
\cite{PoWi90} is based on the same idea, and is quite similar to the
construction of soliton creation operators suggested by Fr\"ohlich and
Marchetti \cite{FrMa87}. It is convenient to consider the model dual to the
original one \re{ZNC}. As shown in Appendix C, its partition
function has the form:

\beq
\cZ^d = \const \sum_{\dual p (\dual \CK{2})\in \Z}
\, \intinf \dual C \expb{ -\frac{1}{2 \kappa} \|\dd
\dual p \|^2 - \frac{N^2 e^2}{2} \| \dd \dual C + \dual p\|^2}.  \label{ZNCd}
\eeq
The dual model describes the interaction of the integer valued hypergauge
field $\dual p (\dual c_2)$ (antisymmetric rank 2 tensor)
with the real valued gauge field $\dual C(\dual c_3)$; the action is
invariant under the hypergauge transformations:

\beq
 \dual p' = \dual p + \dd \dual r; \ \dual C' = \dual C - \dual r.
\label{Htr}
\eeq
Thus we have three equivalent
representations of the partition function: the original one \re{ZNC}, the
BKT--representation \re{ZNCBKT}, and the dual representation \re{ZNCd}.

        Consider now the Wilson loop $W(\cC) = \expb{i(\dual C,\dual
j_{\cC})}$, where the current $\dual j_{C}(\dual c_3)$ is equal to unity on
the links of the dual lattice which belong to the loop $\cC$, and vanishes on
the other links. $W(\cC)$ is gauge invariant, but not hypergauge invariant.
In order to make $W(\cC)$ hypergauge invariant, we use the analogue
of the Dirac procedure, namely we surround the loop by the cloud of the
hypergauge bosons:

\beq
        U_\cC = W(\cC) \cdot \expb{i (\dual D_\cC, \dual p)},     \label{UC}
\eeq
where $D_\cC$ satisfies the equation: $\delta^{(3)} \dual D_\cC = \dual
j_\cC$.
Since the operator of the creation of the string should act at a
definite time slice, we use the three-dimensional operator of the
codifferentiation $\delta^{(3)}$, and the loop $\cC$ belongs to the
considered time slice. It is easy to see that the operator
\re{UC} is invariant under the hypergauge transformations \re{Htr}: $U_\cC'
= U_\cC \expb{ - i (\dual r, \dual j_\cC) + i (\delta  \dual D_\cC, \dual
r)} = U_\cC$. The quantum average of this operator in the BKT
representation,

\beq
    <U_\cC> = \frac{1}{\cZ^{BKT}}
       \sum_{\stackrel{\scriptstyle \dual \sigma(\dual\CK{2})\in \Z}
        {\delta \dual \sigma=\dual j_{\cC}}}
        \expb{- 2\pi^2\kappa \left( (\dual
\sigma - \dual D_\cC),(\Delta + m^2)^{-1} (\dual \sigma - \dual D_\cC)\right)}
\label{UCBKT}
\eeq
shows, that this is indeed string creation operator, since the above sum
is taken over all closed world sheets of the strings, and over all world
sheets spanned on the contour \cC.

         Performing the inverse duality transformation of the average
value of the operator \re{UC},
we get the expectation value of the string creation operator in terms of
the original fields:

\beq \label{ZNCUC} <U_\cC> =
  \frac{1}{{\cZ}}\nsum{l}{1}\intinf A\intpi\varphi \\
  \expb{S_l(A,\dd\varphi -
  2\pi\delta\Delta^{-1}(D_{\cC}-\rho_{\cC})}.
\eeq
Here the integer valued field $\rho_{\cC}$ satisfies the
equation $\delta^{(3)}(\dual D_{\cC} - \dual\rho_{\cC})=0 $;
$\dual \rho_{\cC}$ is the analog of the (invisible) Dirac
string. The Dirac string connected to the monopole is a one-dimensional
object, while $\dual\rho_{\cC}$, being defined on the plaquettes, is a
two-dimensional one. The invisibility of $\rho$ follows
from the invariance of the $<U_\cC>$ given by \re{ZNCUC}
under the deformations of the ``Dirac sheet'': $\rho' =
\rho + \dd \xi$.

\section{Linking of Strings World Sheets and Particle World Trajectories}

        The approach considered here allows us to understand more clearly
the four-dimensional analogue of the Aharonov -- Bohm effect, discussed in
\cite{AlWi89,AlMaWi90,PrKr90}. Let us
calculate the quantum average of the Wilson loop for the charge $Me$,
$W_M(\cC) = \expb{i M (A,j_\cC)}$, in the BKT representation. Repeating all
steps which transform \re{ZNC} into \re{ZNCBKT} we get:

\beqn
<W_M(\cC)> = \frac{1}{\cZ^{BKT}}\nddsum{\sigma}{2} \exp\left\{-
2\pi^2\kappa (\dual\sigma,(\Delta + m^2)^{-1}\dual \sigma) \right.
\makebox[5.4em]{}\label{WNC} \\
 \left.- \frac{M^2 e^2}{2}( j_\cC,(\Delta + m^2)^{-1} j_\cC) - 2 \pi i
 \frac{M}{N} (j_\cC,(\Delta + m^2)^{-1}\delta \sigma) + 2 \pi i \frac{M}{N}
 \LL(\sigma,j_\cC)\right\}.  \nonumber
\eeqn
The first three terms in the exponent describe the short--range (Yukawa)
interactions: surface -- surface, current -- current and current -- surface.
In spite of the gauge field acquiring the mass $m =N \kappa^\half e$, there
is {\it long--range} interaction of geometrical nature, described
by the last term in the exponent $\LL (\sigma,j_\cC)$, $\LL$ being the
four--dimensional analogue of the Gauss linking number for loops in three
dimensions, \ie\ the linking number of surfaces defined by
$\{\sigma\}$ and loop defined by $j_\cC$. The explicit expression for $\LL$
is:

\beq
       \LL = (\dual j_\cC, {\Delta}^{-1} \dd \dual \sigma) =
       (\dual j_\cC, \dual n)
\label{L}
\eeq
where $\dual n$ is an integer valued 3-form which is the solution of the
equation: $\delta \dual n = \dual \sigma$. It is clear now that
$\LL$ is equal to
the number of points at which the loop $j_\cC$ intersects the
three--dimensional volume $\dual n$ bounded by the closed surface defined by
$\dual \sigma(\dual c_2)$. The elements of the surface $\dual \sigma$ may
carry any integer number, so that any intersection point may contribute an
integer into $\LL$.
Therefore $\LL$ is the linking number of
the world sheet of the strings and the current $j_\cC$ which define Wilson's
loop $W_M(\cC)$. The reason for the long--range interaction is that the
charges $e,\ 2e, \ \ldots (N-1)e$ cannot be completely screened by the
condensate of the field of charge $Ne$; if $M/N$ is integer, then the
screening is complete and there are no long--range forces.
The
long--range particle--particle interaction may appear in that phase of the
theory where the condensate of strings exists, and $\LL(\sigma,j_\cC)$ does
not vanish for large Wilson loops. The dynamical properties of quantum
Nielsen--Olesen strings are discussed in \cite{BPPPW93}.

        Another interesting operator, which can be calculated
{\it exactly}, was suggested in \cite{PrKr90}; this operator is the
product of the Wilson loop $W_M(\cC)$ and the operator suggested in
\cite{KrWi89}, which creates the world sheet of the string on the closed
surface $\Sigma$:

\beq
    F_N(\Sigma) = \nsum{l}{1}\expb{-S_l\left(A - \frac{2\pi
k}{N},\dd\varphi\right) + S_l(A,\dd\varphi)},
\eeq
where $k$ defines the surface $\Sigma$ on the dual lattice: $\delta \dual k
= \delta_\Sigma$; $\delta_\Sigma$ is the lattice $\delta$--function which is
equal to unity on the plaquettes of the dual lattice belonging to the
surface $\Sigma$, and $\delta_\Sigma$ vanishes on all other plaquettes.
We can change the integration variable: $A \rightarrow A + \frac{2\pi
k}{N}$ therefore $<F_N(\Sigma)>=1$. The operator which has a nontrivial
expectation value has the form \cite{PrKr90}:

\beq
        A_{NM}(\Sigma,\cC) = F_N(\Sigma) \cdot \frac{W_M(\cC)}{<W_M(\cC)>}.
\eeq
Performing the same steps which lead to \re{WNC} we get:

\beq
        <A_{NM}(\Sigma,\cC)> = e^{2 \pi i \frac{M}{N} \LL(\Sigma,\cC)}.
\eeq
The meaning of this result is very simple. If the surface $\Sigma$ lies in a
given time slice, then $F(\Sigma) =
\expb{\frac{2\pi i}{Ne}Q_{\Sigma}}$ (see \cite{KrWi89,PrKr90}), where
$Q_\Sigma$ is the total charge inside the volume bounded by the surface
$\Sigma$; if $\LL(\Sigma,\cC) = n$ then there is charge $Mne$ in the
volume bounded by $\Sigma$.

\section{Conclusions; Acknowledgments}

        We have shown that the partition function of Abelian Higgs theory
can be represented as the sum over world sheets of the Nielsen--Olesen
strings; the dynamics of these strings (\eg\ scattering and decay
properties) can be studied by means of string creation and annihilation
operators. Similar analysis can be carried out in the case of a compact
gauge field \cite{BPPPW93}. In this case, the monopoles are present in the
theory, and strings can be open carrying monopole and antimonopole on
their ends. String--like excitations exist also in the scalar theory without
gauge field (global strings). Our formulas \re{ZNC}--\re{ZNCUC} are valid in
this case ($e=0, \, m = 0$). Now the topological interaction is absent, but
there exists the long--range interaction $\Delta^{-1}$ which is due to the
cloud of Goldstone bosons surrounding the string. Our formulas are also
valid for the three--dimensional case. For example for $D=3$, $e=0, \, m =
0$, we have a theory of vortices in the two--dimensional superfluid.
The dual formulation of the continuum Abelian--Higgs
theory has been recently discussed in \cite{Lee93}.

The work of MIP and MAZ has been partially supported by a
grant of the American Physical Society. The authors are grateful to
M.Minchev and to T.L.Ivanenko for interesting discussions. MIP expresses his
thanks to HLRZ in J\"{u}lich for hospitality.

\Appendix{ }
Here we briefly summarize the main notions from the
theory of differential forms on the lattice \cite{BeJo82}.
The advantages of the calculus of
differential forms consists in the general character of the expressions
obtained. Most of the transformations depend neither on the space--time
dimension, nor on the rank of the fields. With minor modifications the
transformations are valid for lattices of any form (triangular, hypercubic,
random, \etc). A differential form of rank $k$ on the lattice is a function
$\phi_{k}$ defined on $k$-dimensional
cells $c_k$ of the lattice, \eg\ the scalar (gauge) field is a 0--form
(1--form). The exterior differential operator
{\it d} is defined as follows:

\beq
(\dd \phi ) (c_{k+1}) =\sum_{\CK{k} \in \partial\CK{k+1}} \phi(c_{k}).
\label{def-dd}
\eeq
Here $\partial c_{k}$ is the oriented boundary of the $k$-cell
$c_{k}$. Thus the operator {\it d} increases the rank of the form by unity;
$\dd \varphi$ is the link
variable constructed, as usual, in terms of the site angles $\varphi$, and
$\dd A$ is the plaquette variable constructed from the link variables $A$.
The scalar product is defined in the standard way:
if $\varphi$ and $\psi$ are $k$-forms, then
$(\varphi,\psi)=\sum_{c_k}\varphi(c_k)\psi(c_k)$, where $\sum_{c_k}$ is the
sum over all sells $c_k$.
To any $k$--form on the $D$--dimensional lattice there
corresponds a $(D-k)$--form $\dual\Phi(\dual c_k)$ on the dual lattice,
$\dual c_k$ being the $(D-k)$--dimensional cell on the dual lattice. The
codifferential $\delta=\dual \dd \dual$ satisfies the partial
integration rule: $(\varphi,\delta\psi)=(\dd\varphi,\psi)$.
Note that $\delta \Phi(c_k)$ is a $(k-1)$--form and
$\delta \Phi(c_0) = 0$. The norm is defined by: $\|a\|^2=(a,a)$; therefore,
$\|\dd\varphi+2\pi l\|^2$ in \re{ZNCS} implies summation over all links.
$\nsum{l}{1}$ denotes the sum over all configurations of the integers $l$
attached to the links $c_1$. The action \re{ZNCS} is invariant under the
gauge transformations $A' = A + \dd \alpha$, $\varphi' = \varphi + \alpha$
due to the well known property $\dd^2 = \delta^2 = 0$. The
lattice Laplacian is defined by: $\Delta = \dd\delta + \delta\dd$.

\appendixn{ }

To derive eq.\re{ZNCBKT} we first change the summation variable in \re{ZNC}
(see \cite{Sei82}, Part 1, Chapter 4): $\displaystyle{\nsum{l}{1} =
\sum_{\stackrel{\scriptstyle \sigma(\CK{2}) \in \Z} {\dd \sigma=0}}
\nsum{q}{0}}$, here $l = m[\sigma] + \dd q$ and $m[\sigma]$ is a particular
solution of the equation $\dd m[\sigma] = \sigma$. Using the Hodge
decomposition $m[\sigma] = \delta \Delta^{-1} \sigma + \dd
\Delta^{-1} \delta m[\sigma]$ we introduce the noncompact field $\Phi =
\varphi + 2 \pi (\Delta^{-1}\delta m[\sigma] +q), \ \displaystyle{
\nsum{q}{0}} \intpi \varphi = \intinf \Phi$, and we get:

\beq
   \cZ = \intinf A {\cal D}\Phi
        \sum_{\stackrel{\scriptstyle \sigma(\CK{2}) \in \Z} {\dd \sigma=0}}
\expb{- \frac{1}{2 e^2} \|\dd A\|^2
- \frac{\kappa}{2} \|\dd\Phi + 2 \pi \Delta^{-1}\delta \sigma - N A\|^2}.
\eeq
After fixing the gauge $\dd\Phi = 0$, the Gaussian integral over $A$ can be
easily calculated, and thus we get \re{ZNCBKT}.

\appendixn{ }

        To perform the duality transformation of the original theory defined
by the partition function \re{ZNC}, we change the dynamical
variables, introducing the unity: $1 = \intinf F \cD B \,\delta(B - \dd
\varphi - 2 \pi l + N A)\,\delta(F - \dd A)$ into the integral in \re{ZNC}.
On application of the Poisson summation formula $2\pi \sum_l \delta(x - 2
\pi l) = \sum_l \expb{i l x}$ and the standard representation $\delta (F -
\dd A) = \mbox{\rm const} \int \cD G \expb{i (G,(F - \dd A))}$, the
integrals over $F$ and $B$ become Gaussian, the integrals over $\varphi$ and
$A$ give the restrictions $\delta l = 0$ and $\delta G = l N$, which are
solved by the introduction of new variables $p$ and $C$: $l = \dual
\dd \dual p(\dual c_2)$, $G = N \dual \dd \dual C(\dual c_3) + N p$.
Integrating over $F$ and $B$ we finally get
\re{ZNCd}.


\end{document}